\documentclass[journal]{IEEEtran}
\usepackage{amsmath,amssymb}
\usepackage{subfigure}
\usepackage{cite,balance}
\usepackage{algorithm}
\usepackage{accents}
\usepackage{amsthm}
\usepackage{bm}
\usepackage{url}
\usepackage{algorithmic}
\usepackage{multirow}
\usepackage{enumerate}
\usepackage{cases}
\usepackage{stfloats}
\usepackage{dsfont}
\usepackage{color,soul}
\usepackage{amsfonts}
\usepackage{tcolorbox}
\usepackage{cite,graphicx,amsmath,amssymb}
\usepackage{subfigure}
\usepackage{fancyhdr}
\usepackage{hhline}
\usepackage{array,color}
\usepackage{amsmath}
\usepackage{amsthm}

\usepackage{array}  

\usepackage{booktabs}  

\include{header}

\IEEEoverridecommandlockouts

\begin{document}

\title{From Ground to Sky: Architectures, Applications, and Challenges Shaping Low-Altitude Wireless Networks}

\author{Weijie Yuan,~\IEEEmembership{Senior Member,~IEEE,} Yuanhao Cui,~\IEEEmembership{Member,~IEEE,} Jiacheng Wang,~\IEEEmembership{Member,~IEEE,} \\Fan Liu,~\IEEEmembership{Senior Member,~IEEE,} Lin Zhou,~\IEEEmembership{Senior Member,~IEEE,} Geng Sun,~\IEEEmembership{Senior Member,~IEEE,} \\Tao Xiang,~\IEEEmembership{Senior Member,~IEEE,} Jie Xu,~\IEEEmembership{Fellow,~IEEE,} Shi Jin,~\IEEEmembership{Fellow,~IEEE,} Sinem Coleri,~\IEEEmembership{Fellow,~IEEE},  \\Sumei Sun,~\IEEEmembership{Fellow,~IEEE,} Shiwen Mao,~\IEEEmembership{Fellow,~IEEE,} Dong In Kim,~\IEEEmembership{Life Fellow,~IEEE,}  Abbas Jamalipour,~\IEEEmembership{Fellow,~IEEE,} Mohamed-Slim Alouini,~\IEEEmembership{Fellow,~IEEE,} and Xuemin Shen,~\IEEEmembership{Fellow,~IEEE}\\
\emph{(Invited Paper)}

\thanks{W. Yuan, Y. Cui, and L. Zhou are with the Southern University of Science and Technology, Shenzhen 518055, China (e-mail:yuanwj@sustech.edu.cn; yuanhao.cui@bupt.edu.cn; zhoul9@sustech.edu.cn)}
\thanks{
F. Liu and S. Jin are with the Southeast University, Nanjing 210096, China (e-mail: fan.liu@seu.edu.cn; jinshi@seu.edu.cn)}
\thanks{
J. Wang is with the Nanyang Technological University, Singapore 639798 (e-mail: jiacheng.wang@ntu.edu.sg)}
\thanks{G. Sun is with the Jilin University, Changchun 130012, China (e-mail: sungeng@jlu.edu.cn)}
\thanks{T. Xiang is with the Chongqing University, Chongqing 400044, China (e-mail: txiang@cqu.edu.cn)}
\thanks{J. Xu is with Shenzhen Future Network of Intelligence Institute and The Chinese University of Hong Kong, Shenzhen 518172, China (e-mail: xujie@cuhk.edu.cn)}
\thanks{S. Coleri is with the Koc University, 34450 Istanbul, Turkey (e-mail: scoleri@ku.edu.tr)}
\thanks{S. Sun is with the Institute for Infocomm Research, Agency for Science, Technology and Research (A*STAR), Singapore 138632 (e-mail: sunsm@i2r.a-star.edu.sg)}
\thanks{S. Mao is with the Auburn University, Auburn, AL 36849 USA (e-mail: smao@ieee.org)}
\thanks{A. Jamalipour is with the University of Sydney, Sydney, NSW 2006, Australia and Tohoku University, Japan (e-mail: a.jamalipour@ieee.org)}
\thanks{D. I. Kim is with the Sungkyunkwan University, Suwon 16419, South Korea (e-mail: dongin@skku.edu)}
\thanks{M.-S. Alouini is with the King Abdullah University of Science and Technology, Thuwal 23955, Saudi Arabia (e-mail: slim.alouini@kaust.edu.sa)}
\thanks{X. Shen is with the University of Waterloo, Waterloo, ON N2L 3G1, Canada (e-mail: sshen@uwaterloo.ca).
}
}

\maketitle

\setlength\abovedisplayskip{2pt}
\setlength\belowdisplayskip{2pt}

\vspace{-1.5cm}

\begin{abstract}
In this article, we introduce a novel low-altitude wireless network (LAWN), which is a reconfigurable, three-dimensional (3D) layered architecture. In particular, the LAWN integrates connectivity, sensing, control, and intelligence across aerial and terrestrial nodes that enable seamless operation in complex, dynamic, and mission-critical environments. Different from the conventional aerial communication systems, LAWN’s distinctive feature is its tight integration of functional planes in which multiple functionalities continually reshape themselves to operate safely and efficiently in the low-altitude sky. With the LAWN, we discuss several enabling technologies, such as integrated sensing and communication (ISAC), semantic communication, and fully-actuated control systems. Finally, we identify potential applications and key cross-layer challenges. This article offers a comprehensive roadmap for future research and development in the low-altitude airspace.
\end{abstract}

\section{Introduction}\label{sec:intro}

\begin{figure}[t]
\centering
\includegraphics[width=.5\textwidth]{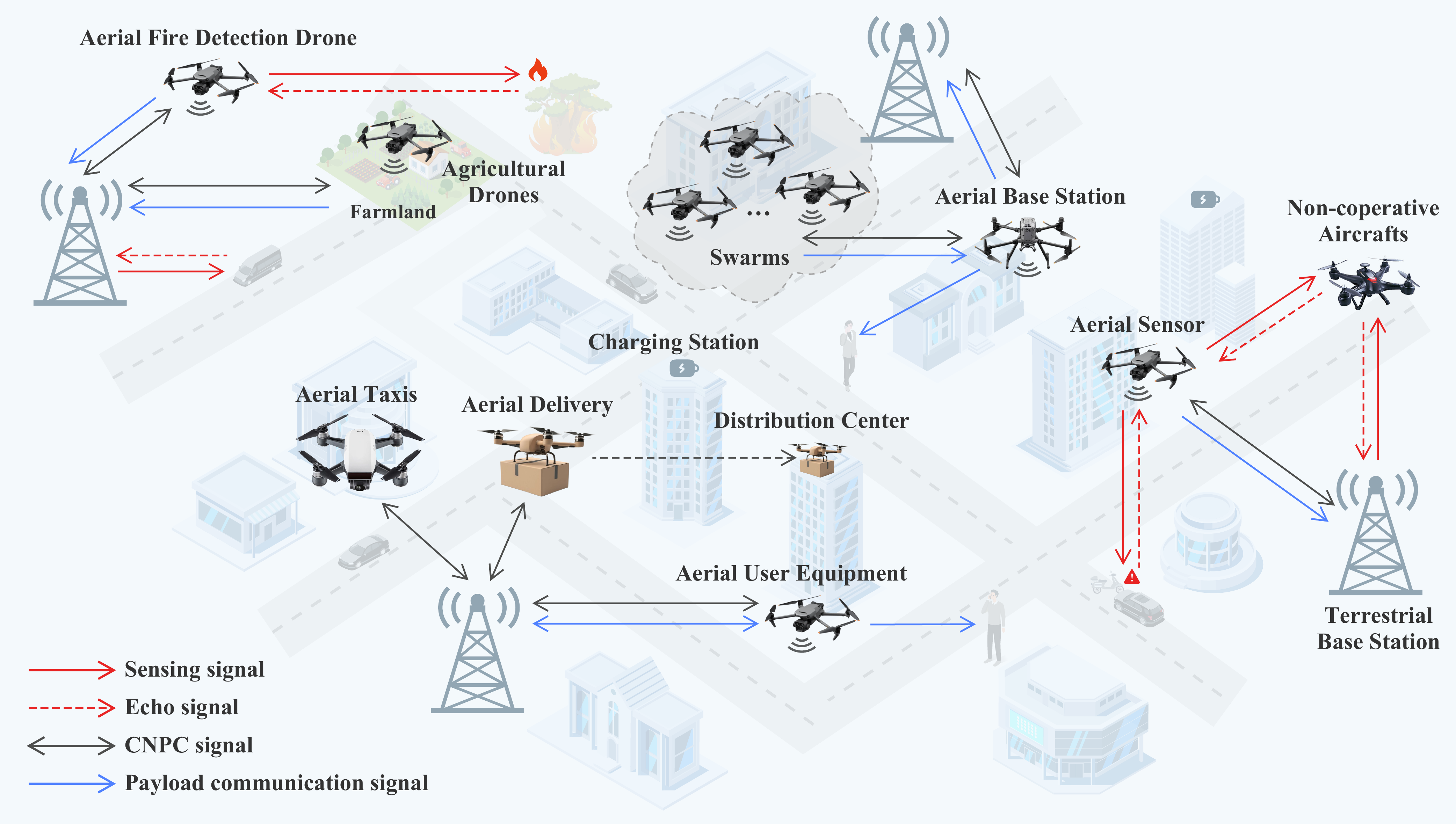}
\caption{The low-altitude wireless network (LAWN) framework integrates a diverse ecosystem of aerial platforms operating below 3, 000 meters to enable a multi-functional, reconfigurable 3D network. The control and non-payload communication (CNPC) links ensure reliable control, the payload communication signals handle the transmission of mission-specific data, and the sensing signals enable active environmental perception.}
\label{scenario}
\end{figure}

The low-altitude airspace, roughly below $3,000$ m above ground level, has rapidly evolved into a third dimension for digital infrastructure supporting logistics, agriculture, and public safety. It is projected that by 2030, the commercial drone sector will have a value of USD $\$260$ billion, four times higher than today's state\cite{dronemarket}. In parallel, regulators are clearing the skies: the U.S. Federal Aviation Administration (FAA)’s Beyond-Visual-Line-of-Sight rule-making committee issued final recommendations in 2024 that pave the way for routine long-range flights without direct pilot oversight, while Europe’s U-space framework harmonized cooperative traffic management at $120$ m above ground level (AGL). At the same time, the civil aviation administration of China (CAAC) made advanced air mobility and “low-altitude economy” a national plan, charting a phased roadmap for exploiting low-altitude airspace. The convergence of market growth, permissive regulation, and maturing autonomy is giving rise to the low-altitude wireless network (LAWN). By eliminating the constraint of fixed ground stations and enabling rapid topology reconfiguration, LAWNs promise various capabilities for precision agriculture, distributed aerial sensing, swarm-based disaster response, and the broader low-altitude economy, as shown in Fig. \ref{scenario}.

While aerial communications have been investigated extensively in the last decade, a LAWN is more than the concept of ``cellular-in-the-sky” \cite{zeng2019accessing}. 
A LAWN is a flexible three-dimensional web that integrates connectivity, real-time sensing, and closed-loop control across multiple drones and terrestrial nodes. Unlike terrestrial or high-altitude systems, the operational environment of a LAWN is relatively densely cluttered (e.g., up to $100$ aerial drones per km$^2$) and rapidly evolving. For instance, drones may execute aggressive maneuvers such as climbing, hovering, and banking at velocities up to 30 m/s, while line-of-sight (LoS) drone-to-drone or drone-to-ground links can quickly turn into non-line-of-sight (NLoS) behind buildings. As a result, signal paths, interference patterns, and safety envelopes may change within a short duration. Meanwhile, flight control often demands end-to-end latency below $10$ ms with 99.999\% reliability to ensure stability. Meeting these coupled requirements requires a LAWN to be elastic in both topology and function. Different drones within range should autonomously share information. The radio waveform carrying data may also be exploited for sensing so that a drone can ``see'' obstacles while it communicates. The controllers should reroute or re-plan swarm formations to combat environmental dynamics, while coordinating intelligence \& computing resources across drones and edge servers to balance performance and battery consumption. In this sense, a LAWN is inherently cross-disciplinary, where communication, sensing, control, and intelligence continually reshape each other to sustain safe and efficient operation in the low-altitude sky.

Despite the great potential of LAWN for the 6G ecosystem, most prior works address isolated topics such as drone-to-ground communication or trajectory design. This article fills this gap by offering a holistic tutorial perspective on what a LAWN is, why it matters, and how it can be realized. In particular, we commence from LAWN architectures by presenting its functional planes, namely data transmission, control, sensing, and intelligence\&computing, clarifying where it diverges from traditional aerial communications or drone-assisted mesh networks. Next, we highlight enabling technologies such as integrated sensing and communication (ISAC) and artificial intelligence (AI), together with representative applications, e.g., precision agriculture. We then discuss a case study of swarm coordination in post-disaster scenarios. To promote the realization of LAWNs, we summarize key challenges including 3D spectrum management, cyber-physical security, and regulation. Finally, we draw conclusions and outline research directions. We hope that this paper will serve as a tutorial for researchers in this area and inspire low-altitude innovations.

\begin{figure*}[t]
\centering
\includegraphics[width=1\textwidth]{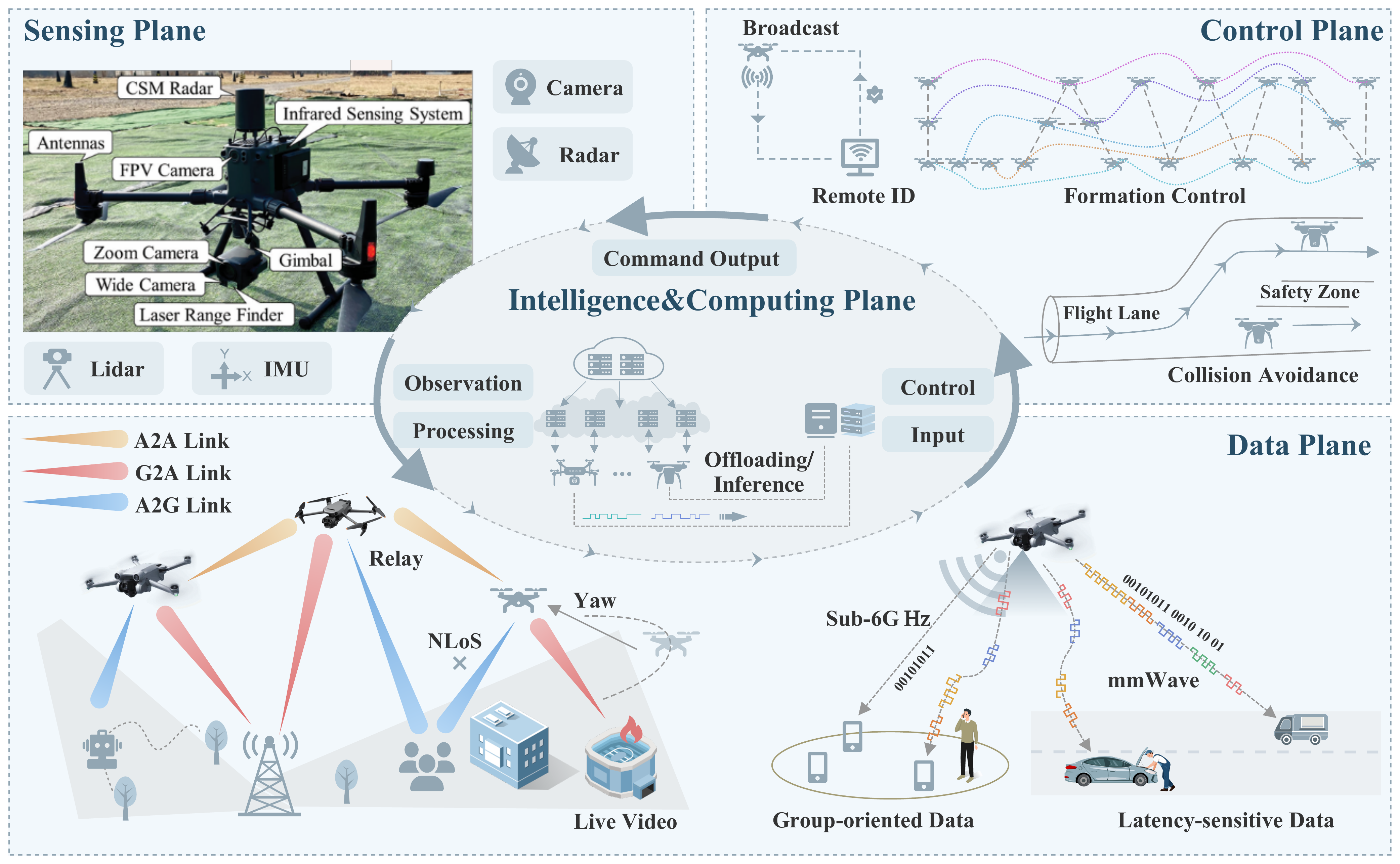}
\caption{The network architecture of a LAWN. There are three tightly coupled functional planes, i.e., the data plane, the control plane, and the sensing plane to provide the capabilities of data transmission, environmental awareness, and flight control. An intelligence\&computing plane further supports distributed inference and hierarchical onboard–edge–cloud processing to enable timely decision-making.}
\label{fig:lawn_layers}
\end{figure*}

\section{LAWN Architectures: A 3-D Web of Connectivity, Sensing, and Control}
The aerial drone-enabled communications can be traced back to the mid-1990s, when drones were considered to serve as platforms for a high capacity trunk radio relay\cite{zeng2019accessing}. Although these ``flying microwave towers'' proved the value of airborne connectivity, their large sizes, high costs, and single-purpose payloads limited wider adoption. Since 2010, the rapid advances in lithium batteries, carbon-composite structures, and system-on-chip radios boosted the research on drone communications. The major benefit of drone communications lies in its flexibility for providing reliable LoS links. 
Therefore, most research to date regards drones purely as airborne base stations or relays to optimize spectral efficiency, mitigate ground-user interference, or reduce drone-to-ground path loss. Nevertheless, future low-altitude missions, such as autonomous delivery, cooperative cinematic swarms, and electric vertical take-off and landing (eVTOL) passenger traffic demand much more than data transmission. They demand real-time sensing, distributed decision-making, swarm coordination, and aviation-grade safety assurance. These multi-dimensional requirements motivate the concept of LAWN, which is to be discussed below.

\subsection{What is the LAWN?}
The LAWN refers to a dynamically reconfigurable three-dimensional network architecture that integrates aerial nodes operating at altitudes typically below $3,000$ m and terrestrial nodes to form a unified platform. This architecture allows for high-speed data transmission ($10-100$ Mbps), ultra-reliable low-latency command-and-control ($<10^{-5}$ error rate and $10$ ms latency), and high-precision sensing (centimeter-level accuracy). The LAWN serves multiple functions, including delivery and logistics for commercial applications, as well as search and rescue operations in disaster scenarios. Fig. \ref{fig:lawn_layers} provides a visual overview of the LAWN architecture. At the heart of the LAWN are three tightly coupled functional planes, i.e., the Data Plane, the Control Plane, and the Sensing Plane, as well as an Intelligence\&Computing (I\&C) Plane.

\subsubsection{\textbf{Data Plane}}
{The data plane} aims to provide on-demand communications, which is responsible for transmitting the mission payloads, such as videos, mapping information, and software updates between aerial drones and terrestrial nodes in the LAWN\cite{zeng2019accessing}. To do so, the data plane supports three fundamental links, i.e.,
\begin{itemize}
\item Air-to-Ground (A2G) link: Mission payloads such as high‑quality video or images, as well as other required information must reach ground operators or edge servers in near real time. The A2G link therefore acts as the primary link of the data plane. 
\item Ground-to-Air (G2A) link: It carries uplink information to the drones, typically including navigation maps, software or model updates, and flight plans. Because the uplink information might arrive in bursts, the G2A link is provisioned as a scheduling priority.  
\item Air-to-Air (A2A) link: It enables each drone to share the sensor data as well as the swarm coordination message to neighboring drones. The A2A connectivity transforms individual drones into a flexible and cooperative mesh, extending coverage beyond any single drone and ensuring robustness when executing tasks.
\end{itemize}
Operationally, the data plane decides who connects to whom, on which band, and with what redundancy, while adapting scheduling and retransmission to LoS/NLoS transitions and kHz Doppler. Although LAWN primarily operates within the low-altitude airspace, high-altitude platforms (HAPs) and low-earth orbit (LEO) satellites can serve as supplementary communication nodes. They provide extended backhaul or remote relay capabilities, especially in remote regions lacking terrestrial infrastructure.

Due to the realities of low‑altitude airspace, there are some common impairments across the data plane of LAWN. A drone can yaw tens of degrees per second and cruise at $10–30$ m/s, turning a clear LoS link to a NLoS one. A data plane scheduler is therefore needed to continuously tag each packet burst with LoS/NLoS status, instantaneous path‑loss, and blockage probability \cite{10040750}. If the tag indicates a positive path state, the data keeps flowing; if a building blocks the signal with a high chance, the scheduler may choose to divert the packets to a nearby relay drone or only transmit the small and time‑critical packets. The relative velocities between different nodes introduce the Doppler shift up to kilohertz, depending on the carrier frequency. In such cases, the data plane will allow coding, hybrid automatic repeat request, and packet‑aggregation timers to shrink on fast‑fading channels. Moreover, for the LAWN operating at heterogeneous bands, e.g., mixed sub-6 GHz and mmWave bands, the data plane could dynamically resize channel bands and link transmit powers by performing the make‑before‑break handoffs. 

The data plane also differentiates traffic flows based on their service requirements. For example, the raw sensor data or software update, which we refer to as \emph{elastic and delay‑tolerant data}, can be buffered first and transmitted only when transmission conditions become favorable. The \emph{latency-sensitive data}, such as live video, should stick to the most stable LoS paths, allowing a seamless hand‑over. \emph{Group‑oriented data}, including swarm perception maps, are broadcast through the aerial mesh, avoiding low-efficiency point-to-point transmission. 

\subsubsection{\textbf{Control Plane}}
{The control plane} can be seen as the nervous system of the LAWN, enforcing aviation-grade safety including identity, separation, and contingency actions while enabling coordinated multi-drone maneuvering under stringent latency/reliability and authenticity constraints \cite{tischler2018system}. On the one hand, it guarantees that every drone in the sky is uniquely identified and properly separated, ensuring the safety to fly. On the other hand, the control plane also enables multiple drones to maneuver in a coordinated group manner. Meeting these goals demands stricter service requirements than the data plane, e.g., a latency of $<10$ ms and reliability of $>99.999$\%, as well as cryptographic authenticity for each packet. 

Building on these requirements, the control plane supports several key functionalities for safe and efficient drone operations in LAWNs:
\begin{itemize}
\item Command and Control (C2): The C2 link serves as the primary interface between each drone and its remote operator or onboard autopilot system. It carries essential flight telemetry, such as position, velocity, orientation, and battery status, and receives control inputs including pathway updates, flight mode transitions, and emergency commands (e.g., return-to-home). Given that failure of the C2 link can result in loss of control or even drone crashes, LAWN architectures must be designed with robust redundancy. Specifically, each drone is equipped with a primary control link operating in sub-GHz or L-band frequencies, and a secondary backup link, typically via a satellite-based channel. This ensures continuous control with minimal disruption to low-altitude operations. 
\item Remote Identification (ID): In accordance with the regulations from FAA, all drones must periodically broadcast a standardized set of identity and flight information. This mechanism, known as remote ID, enables real-time tracking of drone activity and supports dynamic airspace management functions. Within the LAWN, the remote ID beacons are integrated in the control plane via the C2 link. To ensure security and integrity, each beacon will be digitally signed, preventing unauthorized manipulation.
\item Collision Avoidance: Traditional aviation tools such as automatic dependent surveillance-broadcast (ADS-B) are not practical in LAWNs due to bandwidth constraints and device size, while global navigation satellite system (GNSS)-based localization alone cannot provide sufficient positioning accuracy. To address this, the control plane of LAWN may integrate coarse GNSS data with high-resolution range and Doppler measurements derived from the sensing plane. Each drone broadcasts its `safety bubble' including its movement vectors to neighboring drones, which receive and fuse the information into local occupancy maps, enabling real-time assessment of potential conflicts.
\item Formation Control: Many LAWN applications, such as cinematic swarms and emergency rescue operations, require multiple drones to fly in tight coordination while maintaining a desired formation. In the control plane, a designated leader broadcasts shared formation waypoints, with each follower drone adjusting its trajectory based on its assigned offset. These drones exchange high-frequency updates with neighboring drones. Time synchronization beacons embedded in the control channel enable all participants to maintain tightly synchronized control loops.
\end{itemize}
Each of these functionalities plays a critical role in enabling safe and efficient drone operations in LAWNs.

\subsubsection{\textbf{Sensing Plane}}
The sensing plane is dedicated to real-time environmental awareness, which enables aerial and terrestrial nodes to collect environmental information, such as obstacle locations, motion patterns, or atmospheric conditions\cite{boubrima2021robust}. This capability is essential for navigating the complex and often cluttered low-altitude environment, where rapid changes, such as the LoS/NLoS transitions due to the presence of buildings, require immediate responses. By providing accurate and timely environmental data, the sensing plane plays a central role in supporting the coordinated functionality of the LAWN. The sensing plane relies on the following key tools and devices:
\begin{itemize}
\item Onboard Sensors: Each drone in the LAWN is typically equipped with a combination of cameras, LiDAR, mmWave radar, and inertial measurement unit (IMU). These can gather high-resolution imagery, depth information, and acoustic data, essential for detecting obstacles and mapping surroundings in real time. In particular, cameras capture visual data for identifying objects, LiDAR uses laser pulses to measure precise distances to obstacles, and mmWave radar can detect objects in low-visibility conditions, such as fog or darkness. IMUs deliver high-frequency measurements of linear acceleration and angular velocity, which are essential for  motion estimation and stabilizing control loops. 
\item Radio Frequency (RF) Sensing: In addition to passive sensors, the LAWN nodes can exploit radar-like capabilities using radio signals to estimate range and Doppler shift of surrounding objects. A commonly used model is
\begin{equation}\label{eq:isac}
r(t)=\sum_{p}\alpha_{p}\,x(t-\tau_{p})e^{j2\pi \nu_{p} t}+w(t),
\end{equation}
where $x(t)$ is a known probing waveform and $(\tau_p,\nu_p)$ correspond to range and radial velocity of the $p$-th target, respectively. RF waveforms measure $\tau_p$ and $\nu_p$ to determine the position and velocity of objects or other nodes, performing effectively in both LoS and NLoS conditions. These estimates complement cameras/LiDAR/IMU and become more powerful when shared over A2A/A2G links for cooperative sensing.
\item Environmental Sensors: Such sensors could collect atmospheric data, such as temperature, humidity, or air quality, to monitor weather conditions or other environmental factors that impact drone operations and mission planning. The collected data can also be aggregated to create comprehensive environmental models. Environmental sensors usually facilitate environmental monitoring applications, such as tracking air pollution or assessing climate impacts in remote areas.
\end{itemize}
Together, these components empower the sensing plane to provide accurate environmental awareness.

\subsubsection{\textbf{Intelligence \& Computing Plane}}
In addition to the three primary functional planes above, the I\&C plane serves as a supporting functional layer within LAWNs, which is dedicated to providing the computational resources and distributed intelligence required for real-time data processing, signal analysis, and decision-making. Its role enhances the primary functions of connectivity, control, and sensing in LAWNs. The I\&C plane is composed of on-board processors, edge servers, and cloud-based infrastructure. Beyond providing computing resources, this plane also hosts distributed inference, prediction, and decision support that directly shapes scheduling and control.

At the node level, on-board processors enable immediate execution of time-sensitive tasks such as obstacle detection and path planning. For more computationally intensive tasks, such as map reconstruction, the I\&C plane could leverage edge offloading, dynamically assigning workloads from resource-constrained drones to nearby edge nodes\cite{kaleem2019uav}. The edge servers, typically deployed at ground base stations, provide substantial resources for data processing. Through dynamic task offloading, the I\&C plane can reduce onboard computational load while meeting latency requirements. Moreover, cloud-based infrastructure provides high-capacity computing, long-term storage, and model training. While cloud nodes are not suitable for latency-critical operations, they are essential for performing analysis and network optimization. By integrating onboard, edge, and cloud resources, the I\&C plane of LAWNs supports flexible placement of both computation and intelligence across the network, improving scalability and resilience in large-scale deployments.

We can see that all four planes are closely interconnected, working together to support the LAWN's adaptive capabilities. For instance, sensing data about a nearby building informs the control plane to reroute flight paths instantly. Meanwhile, the data plane ensures that this sensing data, along with C2 signals, is transmitted efficiently, while maintaining network-wide awareness. The I\&C plane will process raw sensing data into actionable insights, support distributed inference/learning, optimize control algorithms, and balance workloads between drones and edge servers. This tight integration allows the LAWN to respond seamlessly to rapidly changing conditions, supporting applications such as swarm coordination, disaster response, urban surveillance, package delivery, and environmental monitoring.

\subsection{How LAWN Differs from Aerial Communications?}
Traditional aerial communication networks primarily focus on providing connectivity or extending coverage, as seen in concepts such as ``cellular-in-the-sky\cite{zeng2019accessing}.'' Data transmission and control are typically performed through separate networks and sensing is often handled by individual drones or external systems. The limited integration of control and sensing capabilities restricts their suitability for emerging low-altitude applications. 

\begin{table*}[htbp]
\centering
\caption{Comparison Between Legacy Aerial Architectures and LAWN}
\label{tab:architecture_comparison}
\begin{tabular}{p{0.15\linewidth} p{0.25\linewidth} p{0.25\linewidth} p{0.25\linewidth}}
\toprule
\textbf{Aspect} & \textbf{UAV-Assisted Cellular / Aerial BS} & \textbf{FANET / Drone Mesh} & \textbf{LAWN} \\
\midrule
\textbf{Core Perspective} & Aerial nodes act as flying relays or base stations to serve ground users & Aerial nodes act as self-organizing routers to relay data among a swarm & Extends terrestrial connectivity upward to create a service-aware 3D ``skyway'' for aerial agents \\\hline
\textbf{Functional Scope} & Focused purely on data transmission with separate control and sensing layers & Focused on decentralized data routing while control and sensing are handled locally per drone & Integrated data, control, and sensing planes with an Intelligence \& Computing (I\&C) plane \\\hline
\textbf{Network Structure} & Fixed infrastructure anchors or predefined relay paths & Ad-hoc, decentralized node-to-node topology & Dynamically reconfigurable 3D mesh of tightly coupled aerial and terrestrial nodes \\\hline
\textbf{Sensing Capability} & Relies entirely on external systems & Local sensing by individual drones & Cooperative sensing with shared perception across nodes for enhanced situational awareness \\\hline
\textbf{Computing Resources} & Centralized at the ground core network & Onboard-only processing and limited scalability for heavy tasks & Hierarchical computing (onboard, edge, and cloud) for real-time processing and decision-making \\
\bottomrule
\end{tabular}
\end{table*}

{In contrast, the LAWN introduces a paradigm shift by redefining the network around the terrestrial infrastructure, extending its functionality into the low-altitude space through a reconfigurable, intelligent, and mission-aware network ``skyway''. Conventional UAV-assisted cellular networks and aerial base stations primarily act as flying relays to extend terrestrial coverage, focusing almost exclusively on the data plane. On the other hand, Flying Ad-Hoc Networks (FANETs) or drone meshes excel at air-to-air decentralized routing but typically treat flight control and environmental perception as isolated, standalone tasks. Instead of simply using drones as communication tools, LAWN transcends both by treating aerial and terrestrial nodes as jointly organized components of a cyber-physical system. The architecture tightly integrates three primary functional planes, i.e., data, control, and sensing, augmented by an I\&C plane. Dynamic reconfigurability of both aerial and terrestrial nodes sets LAWN apart from fixed infrastructure configurations, ensuring robustness across diverse, dynamic scenarios. A defining advantage of LAWN is its cooperative sensing capability, allowing aerial and ground nodes to share environmental data in real time. This shared perception supports safety-critical services such as multi-drone formation flying, dynamic obstacle avoidance, and urban air traffic management. Additionally, LAWN’s I\&C plane provides distributed computational resources for real-time processing and decision-making. This includes on-board processing for immediate tasks, edge computing for latency-sensitive operations, and cloud-based infrastructure for intensive computations and long-term data analysis. Such a hierarchical framework is rarely embedded in traditional systems. In Table \ref{tab:architecture_comparison} we summarize the comparison between LAWN, UAV-assisted cellular networks, and FANETs.}


\subsection{Standardization for LAWN-NTN Integration}
To realize the full potential of LAWN and ensure seamless integration with the broader telecommunication ecosystem, aligning LAWN with emerging standards is crucial. The 3rd Generation Partnership Project (3GPP) has been proactively evolving its specifications to support NTN, creating an essential bridge towards the forthcoming 6G infrastructure.

The 3GPP has recently taken initial steps to accommodate aerial and NTN in its standards. Release 17’s scope of NTN encompassed both satellites and airborne platforms, identifying aerial drones and HAPs as special use cases within the NTN framework \cite{saad2024non}. In parallel, the system architecture work in Release 17 defined new functions to support aerial drone identification, authentication, and authorization over cellular networks. Building on this foundation, Release 18 expanded NTN capabilities to improve coverage and performance and to enable closer integration with terrestrial networks. On the aerial drone side, Release 18 introduced explicit new radio (NR) enhancements to better accommodate drones as network clients \cite{3gpp2024service}. In addition, the new ``Mobile IAB” work item in Release 18 defined procedures for integrated access and backhaul (IAB) nodes mounted on moving vehicles, such as drones. This allows a drone equipped with a 5G base station to dynamically provide coverage to users in a cell, effectively creating a low-altitude cell site that can relocate as needed. 

Despite this progress, current 3GPP specifications remain largely data-plane centric and do not yet provide native support for LAWN’s integrated control, sensing, and I\&C planes. In particular, standard primitives for safety-critical C2 reliability and redundancy, cooperative perception message exchange and trust, and task/model continuity across edge nodes are still missing or only loosely defined, which limits LAWN from operating as a fully integrated cyber-physical platform.

\begin{figure*}[t]
\centering
\includegraphics[width=1\textwidth]{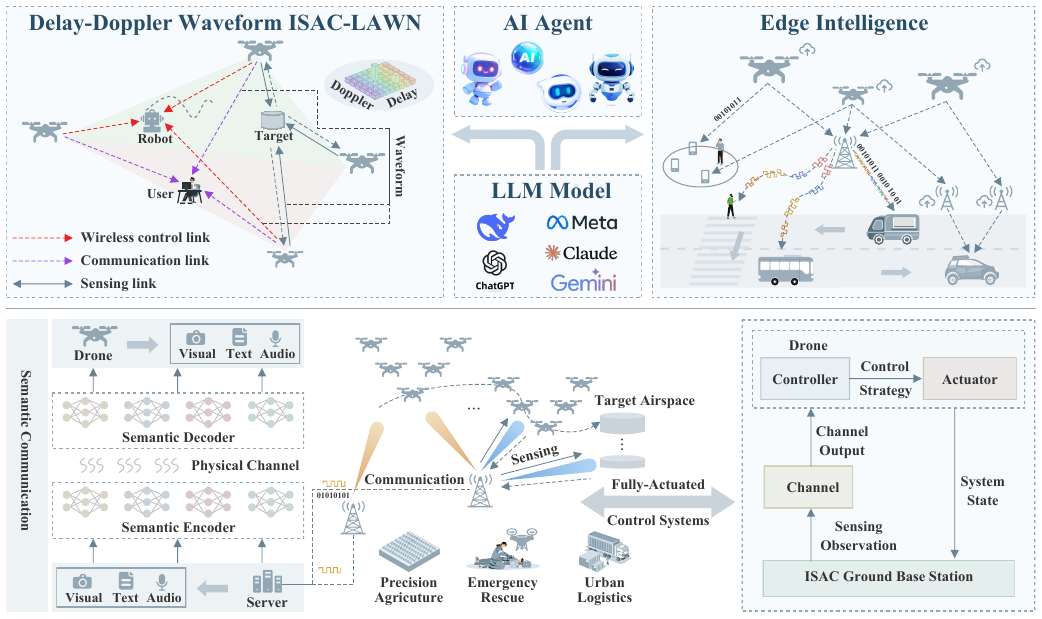}
\caption{Enabling technologies for LAWNs. \textbf{Integrated sensing and communications} merges data delivery with environmental awareness. \textbf{Delay–Doppler waveforms} offer high-mobility robustness. \textbf{Semantic communication} compresses multimodal sensor data. \textbf{Edge intelligence and large language models} distribute perception, prediction, and symbolic reasoning. \textbf{Fully-actuated control systems} close the sensing–control loop through wireless control links.}
\label{fig:enabling}
\end{figure*}

\section{Enabling Technologies and Potential Applications}

The realization of LAWNs relies on a diverse set of enabling technologies spanning physical-layer waveform design, intelligent control, and network-level intelligence, as illustrated in Fig. \ref{fig:enabling}. While the architecture of a LAWN is detailed in Sec. II, advanced tools and protocols are essential to unlock its full potential. 

\subsection{Enabling Technologies}
The realization of LAWNs relies on a diverse set of enabling technologies, spanning physical-layer waveform design, integrated sensing, intelligent control, and network-level intelligence, as illustrated in Fig.~\ref{fig:enabling}. While Sec.~II outlines the functional-plane architecture, the following tools and protocols are essential for achieving mission-grade reliability, situational awareness, and agile autonomy under fast topology changes and severe resource constraints.

\subsubsection{Integrated Sensing and Communications (ISAC)}
ISAC is a transformative technology that enables wireless communication and environmental sensing using the same signaling, hardware platform, and radio resources~\cite{liu2022integrated}. In a LAWN, drones must continuously exchange information while simultaneously perceiving a rapidly changing environment. Therefore, ISAC offers an efficient path to reduce spectrum usage and hardware complexity. For example, a drone transmitting to a ground node can also process reflected echoes to detect obstacles or track other aerial agents, without switching to a dedicated radar or LiDAR module. Unlike classic RF sensing using a probing signal $x(t)$ in \eqref{eq:isac}, ISAC also conveys data information in the transmitted signal $x(t)$. The received echoes $r(t)$ can be processed using classical radar operations such as matched filtering and beamforming, followed by detection/estimation. Moreover, with multiple drones acting as distributed sensors, a LAWN can form a virtual large aperture, improving angular resolution and imaging capability. ISAC can further assist channel state (CSI) acquisition through sensing-assisted approaches, reducing the reliance on frequent pilot transmission \cite{yuan2021integrated}. {For example, twenty questions estimation frameworks~\cite{zhou2025twentyq} could potentially be leveraged for CSI estimation.} Nevertheless, bringing ISAC into compact aerial platforms poses practical challenges in waveform/antenna co-design, hardware limitations, and interference management.

\subsubsection{Delay-Doppler Waveform}
Delay-Doppler (DD) waveforms, including orthogonal time frequency space (OTFS) and its variants, were proposed in 2017 for reliable data transmission in highly dynamic wireless channels. Unlike conventional time-frequency (TF) domain modulation, OTFS modulates symbols in the two-dimensional DD domain\cite{yuan2023new}. Since the physical world merely experiences a slight change in a short time duration, it is expected that the DD parameters associated with the channel are almost unchanged during a short symbol interval. Relying on the DD channel representation, OTFS converts the time-varying channels into near time-invariant ones, providing significant resilience when the channels suffer from high Doppler shifts and rapid channel fluctuations\cite{wei2021orthogonal}. Even if the drone or the scatterers are moving, their effect is concentrated in a few DD bins over the packet duration, making equalization in the DD domain simpler. Furthermore, DD waveform leverages full path diversity (all reflected delays/Dopplers) to maximize throughput under motion. Importantly, DD signaling carries direct physical meaning: a symbol’s spreading in the TF signal corresponds to probing a set of delays and Dopplers. As a result, a LAWN receiver can extract range- and velocity-related information of reflecting objects from the communication signal itself, making DD waveforms attractive for ISAC. From an implementation perspective, the original OTFS framework, built on 2D TF$\rightarrow$DD transformations, is compatible with existing multi-carrier modems through pre-/post-processing blocks. However, the required 2D transforms may become computationally intensive for large bandwidths or high-resolution DD grids, creating tension between robustness and low-latency operation. To address this, Zak-transform-based realizations can directly map the time-domain signal to the DD domain, bypassing explicit TF-domain processing and enabling more efficient implementations for latency-sensitive LAWN links.

\subsubsection{Semantic Communications}
Semantic communications shift the goal from accurate bit-level recovery to successful delivery of task-relevant meaning, intentionally discarding redundancy that does not affect the downstream task. Rather than reconstructing raw messages with minimal bit error, semantic systems interpret context, intent, and importance, and transmit only representations that materially influence the receiver’s inference or action. {This design can be captured by a task-oriented objective
\begin{align}
&\min \;\mathbb{E}\!\left[\mathcal{L}_{\mathrm{task}}(s,\hat{s})\right]\\
&\text{s.t.}
R \le R_{\max},\;\; \Pr(\mathrm{unsafe})\le \epsilon,\nonumber
\end{align}
where $s$ denotes the underlying task-relevant state, $\hat{s}$ is its reconstruction at the receiver, and $\mathcal{L}_{task}(\cdot)$ measures task degradation such as missed detection cost or control-performance loss. Furthermore, $R$ is the utilized data rate bounded by the maximum channel capacity $R_{max}$, and $\Pr(\text{unsafe})$ is the probability of a safety violation bounded by a stringent threshold $\epsilon$.}  In LAWNs, a semantic-aware node may extract high-level descriptions from raw sensor streams and transmit only essential semantic representations, as in interest-based image-segment transmission~\cite{10475844}. This can significantly reduce bandwidth consumption and energy usage, which is particularly valuable for battery-constrained drones. For example, instead of streaming a full 4K video feed, a drone may transmit a concise semantic report such as ``three pedestrians crossing in two seconds,'' enabling other nodes to react without processing raw imagery. Semantic awareness can also support smarter control and scheduling by prioritizing flows based on task importance. However, semantic compression typically requires onboard or edge intelligence for interpretation, and it raises trust and correctness concerns: in safety-critical missions, incorrect or over-compressed semantic messages may lead to poor decisions\cite{li2021cross}.

\subsubsection{Artificial Intelligence and Large Language Models}
AI is indeed the backbone technology in enabling the autonomy and intelligence of the LAWN \cite{zhao2025generative}. It is closely connected with the capabilities for perception, decision-making, and cooperative behavior across spatially distributed nodes. By deploying lightweight neural networks on drones or edge servers, a wide range of real-time tasks such as channel prediction, trajectory optimization, obstacle classification, and intent estimation can be supported. The drones can be trained to navigate through complex environments or select optimal relay positions through reinforcement learning algorithms. The emergence of large language models (LLMs) has offered a new dimension of cognitive functionality, enabling reasoning, instruction interpretation, and symbolic decision-making. When integrated into LAWNs, LLM can translate natural-language instructions into structured mission parameters, enabling intuitive command interfaces for non-expert users\cite{chang2024survey}. In turn, the mission logs, sensor feeds, or visual data obtained by the LAWN can be summarized as natural-language reports for post-mission review. Furthermore, LLM may assist semantic communications within multi-agent coordination through shared symbolic understanding. Ensuring real-time performance of AI algorithms is crucial. Techniques like model compression, quantization, hardware acceleration (using onboard GPUs or NPUs), and offloading computation to edge servers are all employed to make AI feasible for resource-constrained drones. Moreover, the LAWN should incorporate online learning or periodic retraining to ensure that models stay robust over time.

\subsubsection{Edge Intelligence}
Edge intelligence complements the I\&C plane by enabling distributed inference and fusion close to the operational area. A typical hierarchy assigns time-critical, lightweight functions to drones, latency-sensitive and compute-heavy tasks to edge servers, and long-term analytics/model training to cloud nodes, enabling dynamic task partitioning~\cite{kaleem2019uav}. For example, a drone may keep immediate obstacle avoidance on-board while offloading compute-intensive recognition or multi-view fusion to a nearby edge server. {A simple latency model that highlights the feasibility of offloading is
\begin{equation}
T_{\mathrm{off}}=\frac{b}{R}+\frac{c}{f_{\mathrm{e}}}+T_{\mathrm{return}},
\end{equation}
where $b$ is the transmitted feature/data size, $R$ is the achievable uplink rate, $c$ is the required compute cycles, $f_{\mathrm{e}}$ is the edge processing rate, and $T_{\mathrm{return}}$ accounts for result delivery and protocol overhead. This expression illustrates that offloading is only beneficial when the combined delays of wireless transmission ($b/R$) and edge computation ($c/f_e$) remain strictly lower than the time it would take a drone to process the data locally.} A key capability is edge perception, where multimodal sensing data from multiple drones is aggregated at edge nodes for high-level scene understanding. The edge then distributes compact perception outputs back to the swarm, allowing each drone to benefit from shared situational awareness while limiting onboard processing and inter-drone traffic. A central challenge is 3D mobility, which causes intermittent connectivity and demands seamless task/model continuity across edge nodes during movement and handover.

\subsubsection{Multi-Connectivity and Redundant Networking}
Mission-grade LAWN operation requires negligible performance degradation under blockage, interference, and node failures. Multi-connectivity provides a practical path to reliability by maintaining parallel links, e.g., sub-6~GHz control anchor with mmWave/FR2 data boosting, or terrestrial plus satellite backup\cite{pupiales2021multi}. Combined with packet duplication, opportunistic relaying over the A2A mesh, and lightweight coding strategies for multicast/group-oriented traffic, multi-connectivity can reduce disruption during fast LoS/NLoS transitions and make-before-break handoffs. In addition, redundant networking supports safety-critical traffic isolation, ensuring that C2 and remote-ID signaling remain protected even when mission payload traffic surges.

\subsubsection{Fully-Actuated Control Systems}
Traditional drones often rely on state-space control models that work well for linear systems. However, these approaches may become inadequate in dynamic, uncertain, or time-varying environments. The fully actuated system (FAS) approach offers a new control paradigm that enables precise, robust, and real-time control design for highly nonlinear and time-varying systems\cite{10689714}. The FAS approach constructs a control-oriented model structure that allows the control input to be explicitly solved, bypassing the need for standard state-space variable extensions. By converting the original physical model into a mathematically equivalent fully actuated form, the FAS framework enables direct controller design with arbitrarily assignable system dynamics. In the LAWN, aerial nodes modeled using the FAS framework can be stabilized via linear or nonlinear control laws that directly assign desired performance, even in the presence of modeling uncertainty, communication delays, and external disturbances. Moreover, FAS provides effective control even for systems that are not feedback linearizable, making it highly suitable for uncertain aerial dynamics in low-altitude environment. With its theoretical guarantees, fully-actuated control systems strengthen the autonomy, agility, and safety of LAWNs.

\subsubsection{Wireless Power Transfer}
Energy supply is one of the most critical constraints for LAWNs. Conventional drones require frequent landings for recharging, which limits mission continuity. Wireless power transfer (WPT) can enable in-flight or hovering refueling without physical contact~\cite{xie2021uav}. For far-field WPT using directional microwave/mmWave beams, the harvested energy over a charging interval $T$ can be abstracted as
\begin{equation}
E_{\mathrm{WPT}}=\eta\,P_{\mathrm{t}}\,G_{\mathrm{t}}G_{\mathrm{r}}\!\left(\frac{\lambda}{4\pi d}\right)^{2}T,
\end{equation}
where $\eta$ is the end-to-end power-transfer efficiency, $P_{\mathrm{t}}$ is the transmit power, $G_{\mathrm{t}}$ and $G_{\mathrm{r}}$ are transmit/receive beamforming gains, $\lambda$ is the wavelength, and $d$ is the separation distance. It can be observed that the harvested energy relies on geometry and beam alignment, implying that WPT must be co-designed with routing and formation control, e.g., by guiding drones into favorable charging locations and maintaining beams under mobility.

\subsection{Applications enabled by LAWNs}
LAWNs unlock a broad range of applications across consumer, industrial, and public sectors by combining altitude-segmented air corridors with tightly coupled functional planes. 

Urban logistics is among the most visible application domains for LAWNs, covering parcel delivery, food distribution, and medical supply transport. Vertiports, dynamic geofencing, and integrated traffic management within the control plane keep UAV logistics both efficient and compliant with safety regulations. This is particularly critical for last-mile delivery in dense urban environments and for rapid deployment of essential medical supplies during emergencies \cite{deng2024drone}. 

In rural and regional settings, LAWNs enable large-scale sensing and data collection for crop monitoring, spraying, irrigation planning, and environmental surveillance\cite{yang2017unmanned}. The sensing plane supports multimodal perception to produce high-resolution environmental maps. Data aggregation and processing can be offloaded to edge computing facilities, while long-horizon analytics such as yield prediction or climate impact modeling reside in the cloud. These systems combine centimeter-scale accuracy with edge analytics, providing timely warnings in remote regions. 

In disaster and emergency scenarios, LAWNs offer a way to rapidly restore connectivity and generate situational awareness. When terrestrial infrastructure is damaged, drones can form airborne relay networks to re-establish communication links and provide real-time imagery for search and rescue. The control plane maintains robust C2 connections, while the sensing plane aggregates observations for localizing survivors and identifying critical hazards. Beyond acute emergencies, the same infrastructure can support routine public safety tasks, including traffic monitoring, crime prevention, and firefighting in dense urban areas \cite{perez2023urban}. 

Urban Air Mobility, powered by eVTOL aircraft and other emerging platforms, is one of the most transformative application families enabled by LAWNs. Over the past few years, eVTOL test flights and demonstrations have taken place across North America, Europe, and Asia \cite{cohen2021urban}. Turning these demonstrations into scalable services will require low-altitude corridors with URLLC for safety-of-flight, cooperative sensing for collision avoidance, and seamless integration with UTM/U-space systems. LAWNs provide the necessary infrastructure for these requirements. The control and data planes jointly support fleet coordination, route scheduling, and integration of crewed and uncrewed traffic. The sensing plane supplies shared perception and traffic awareness, while the I\&C plane handles trajectory optimization, demand-responsive routing, and predictive maintenance. 

Infrastructure inspection is another high-impact use case in which LAWNs play a central role. Drones can inspect bridges, pipelines, power lines, offshore wind farms, and other critical assets, often in environments with complex propagation conditions and limited ground access. Field reports indicate that major utilities in California have saved millions of dollars and improved worker safety by switching to drone-based inspections for power lines \cite{pge2025}. 

Beyond mission-critical domains, LAWNs also unlock new applications in entertainment and media \cite{liu2024design}. Coordinated drone swarms have already demonstrated large-scale aerial light shows, surpassing the capabilities of traditional fireworks by offering programmable, environmentally friendly, and reusable visual displays. In addition to these established applications, LAWNs open the door to immersive AR/VR experiences in stadiums and city centers, where synchronized aerial platforms can project visual effects or serve as mobile anchors for mixed-reality overlays. Looking further ahead, LAWNs may support emerging services such as airborne data centers and persistent aerial sensing grids that integrate into smart-city infrastructures.

In summary, the flexibility of LAWNs ensures they can simultaneously support mission-critical services like public safety while also fostering new commercial and consumer ecosystems in the low-altitude airspace.
\subsection{Lessons Learned}
From the applications above, we can find the following insights into the potential of the LAWN:
\begin{itemize}
\item \textbf{Real-Time Data Processing}: Leveraging its integrated data and I\&C planes, the LAWN enables real-time collection and processing of high-resolution data, as seen in precision agriculture, thereby improving resource optimization and decision-making in dynamic environments.
\item \textbf{Dynamic Network Reconfiguration}: Utilizing its reconfigurable 3D architecture, the LAWN supports adaptive operations in urban logistics, allowing seamless adjustments to changing conditions such as traffic or weather.
\item \textbf{Rapid Deployment in Crises}: Through its ad-hoc networking capabilities, the LAWN facilitates robust communication in emergencies, as demonstrated in public safety applications, ensuring operational continuity in disrupted environments.
\item \textbf{Precision Surveillance}: Employing coordinated sensing and control planes, the LAWN enables monitoring for environmental and infrastructure applications, enhancing sustainability through precise fault detection and data collection.
\item \textbf{Cross-Domain Integration}: By integrating with diverse systems like IoT, as shown in emerging applications such as smart cities and wildlife tracking, the LAWN supports versatile, interdisciplinary operations across varied domains.
\end{itemize}

In summary, LAWN’s unified architecture unlocks new applications that demand mobility, adaptability, and edge intelligence. Its ability to simultaneously deliver connectivity, perception, and control makes it a key enabler for the next-generation aerial systems. 

\begin{figure*}[t]
\centering
\includegraphics[width=1\textwidth]{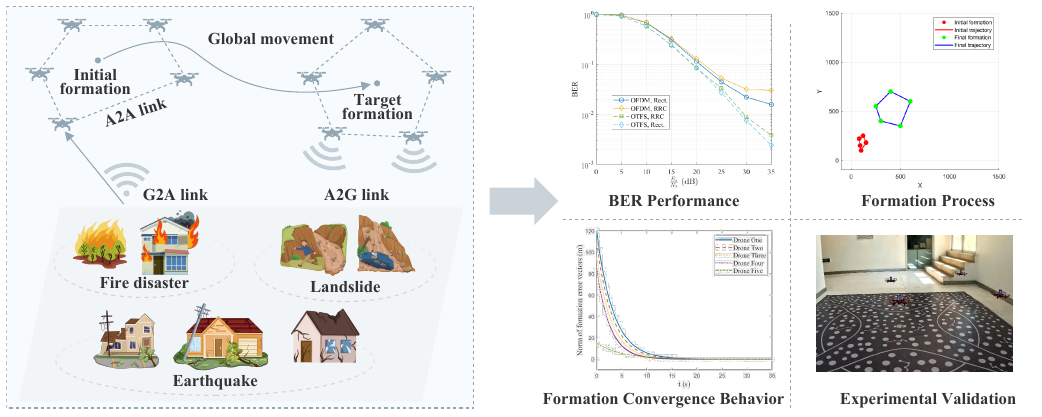}
\caption{Case study of LAWN-assisted swarm coordination in post-disaster scenarios. A ground node disseminates flight/formation updates via G2A links and reliable C2 signaling, collects mission payloads via A2G links, and leverages A2A links among drones for local information sharing and cooperative coordination. 
}
\label{fig:case}
\end{figure*}

\section{Case Study: LAWN-Assisted Swarm Coordination}
Let us consider a post-disaster scenario (e.g. after a hurricane or an earthquake) where first responders deploy a fleet of aerial drones to survey damage and locate survivors\cite{kaleem2019uav}. 

\subsection{System Setup}
In the system setup, we consider that a ground LAWN node (e.g., mobile command center) is equipped with multi-band radio transceivers and edge servers. It maintains a ground link to surviving terrestrial networks, a sub-GHz link for C2 signal, and an LEO satellite link as a backup. The LAWN architecture manages drone operations through four integrated planes: control, data, sensing, and I\&C. Specifically, the control plane employs time-synchronization beacons from the ground node to keep all drones’ control loops aligned. The data plane carries mission payloads, such as high-resolution images, live videos, and sensory readings, via A2G links back to the ground node. Then, updated flight plans or map data are transmitted to the drones via G2A links, as shown in Fig. \ref{fig:case}. The DD waveform is adopted in the data plane to preserve reliability under kHz-level Doppler shifts. 

The sensing plane fuses data from sensors, e.g., camera, LiDAR, and IMU with ISAC echoes, continually mapping the terrain and detecting survivors or obstacles. In a cluttered disaster zone, drones use sensing data to avoid collisions. The sensing plane thus augments the control plane with environmental data, critical when rapid changes, such as LoS/NLoS transitions occur. The I\&C plane allocates heavy data processing tasks across onboard processors as well as the edge servers at the ground LAWN node, which does not slow down the aerial drones. Assuming that the formation task is for the drones to form a desired pattern, the ground LAWN node initially commands a leader drone to an entry point. Then the LAWN control plane ensures that the other drones adjust their flight paths to maintain the desired formation. 

Critically, even if the LAWN ground node loses partial or complete external connectivity, the four planes adapt to ensure continuous and autonomous swarm operation. The A2A links are established between drones to form a cooperative mesh. A2A links share sensed data, e.g., local obstacle maps, among neighboring drones. For instance, if one drone’s view is blocked, it can get surrounding map data via A2A links. Each drone’s onboard computing resources allow lightweight AI agents, enabling the swarm to dynamically reconfigure formations and continue executing mission tasks independently, ensuring robust and resilient operation under all network conditions. 

\subsection{Results}
\subsubsection{Simulations}

{To evaluate the system performance, we simulate a LAWN-assisted swarm coordination scenario featuring a fleet of five drones. The drones' initial positions are uniformly distributed within a $1500 \times 1500 \times 40~\text{m}^3$ volume, and they cruise at velocities ranging from $10$ m/s to $15$ m/s. The swarm's control objective is to converge into and maintain a 40 m-radius pentagon formation.}

{Within the control plane, swarm coordination is driven by a $100$ Hz leader-follower model predictive controller (MPC). To guarantee the necessary sub-10 ms end-to-end control latency, the MPC maintains strict alignment using lightweight synchronization beacons broadcast over a dedicated sub-GHz C2 link. At a transmission rate of $100$ Hz, each beacon's compact 64-byte payload translates to a minimal signaling overhead of just $51.2$ kbps per drone, which is practically negligible given the bandwidth. Additionally, the synchronization protocol strictly bounds the expected jitter to under 100 $\mu$s. This ensures the MPC operates with the exact precision needed for highly dynamic formation maneuvers, successfully avoiding queuing delays and packet collisions. For continuous environmental awareness, the sensing plane leverages ISAC echoes derived from the payload transmissions. Any obstacles detected via these reflections are immediately fed back into the MPC to enable dynamic collision avoidance.}

{The data plane utilizes a DD waveform to handle the mission payload traffic. The communication link is evaluated using a 3D low-altitude channel model featuring three multipath components and a Jakes Doppler distribution, which accurately reflects the maximum relative velocity of $15$ m/s. The DD transmission is configured with a delay-Doppler grid of $M = 64$ and $N = 10$, mapped to a 16-QAM modulation scheme. For benchmarking purposes, an OFDM baseline is implemented with identical bandwidth, subcarrier spacing, and modulation.}

{System reliability is assessed across 1,000 independent Monte Carlo realizations. To ensure a fair comparison, the bit-error-rate (BER) is computed strictly over the extracted payload data bits. The detailed simulation setup is summarized in Table II. As illustrated in Fig. 4, the DD waveform significantly outperforms the OFDM baseline under high-mobility conditions. Furthermore, despite the cluttered environment, the swarm successfully converges from random initial positions to the target pentagon formation within $15$ s on average.}
\begin{table}[htbp]
\centering
\caption{Simulation Parameters for LAWN Swarm Coordination}
\label{tab:sim_setup}
\begin{tabular}{|c|c|}
\hline
\textbf{Parameter Category} & \textbf{Specific Configuration} \\
\hline
\textbf{Swarm \& Kinematics} & 5 drones, $v \in [10, 15]$ m/s \\\hline
\textbf{Control Plane} & 100 Hz , micro-second sync beacons \\\hline
\textbf{Channel Model} & $P=3$ paths, Jakes Doppler spectrum \\\hline
\textbf{Waveform Modulation} &  16-QAM\\\hline
\textbf{OTFS / DD Frame} & $M = 64$ subcarriers, $N = 10$ time slots \\
\hline
\end{tabular}
\end{table}

\subsubsection{Real-world Experiment}
{We also conduct a real-world experiment to validate the system, where the explicitly implemented components include four DJI Tello drones acting as the aerial nodes and a $10\times10$ m$^2$ motion-capture arena providing real-time sensing and localization. Furthermore, a trolley-mounted laptop serves as the ground LAWN node, handling the control and intelligence computations. From arbitrary starts, the drones form a square formation with four translational motions, namely up, forward, down, and backward. The laptop commands these maneuvers to the designated leader. Then, the three followers maintain fixed displacement offsets, allowing the square to translate rigidly while holding formation throughout the motion.}

\begin{table*}[t]
\centering
\caption{Roadmap of Cross-Layer Challenges for LAWNs}
\label{tab:LAWN_roadmap}
\small
\setlength{\tabcolsep}{5pt}
\renewcommand{\arraystretch}{1.5}
\begin{tabular}{|p{0.41\linewidth}|p{0.28\linewidth}|p{0.27\linewidth}|}
\hline
\textbf{Challenge} & \textbf{Potential solutions} & \textbf{Performance metrics} \\
\hline
\textbf{3D spectrum coexistence:} Altitude-/orientation-dependent interference and fast beam re-pointing under mixed frequency bands yield rapidly varying interference footprints
& Dynamic 3D interference modeling, predictive beam/power scheduling, sensing-assisted spectrum awareness
& SINR, outage, handover interruption \\
\hline
\textbf{Time synchronization \& localization integrity:} GNSS degradation, multipath, and clock drift lead to biased locations and timings in drone swarms
& Network-assisted synchronization, integrity monitoring, cooperative localization 
& Synchronization error, integrity-bounded position error/undetected failure, spoofing detection time \\
\hline
\textbf{Low-latency control in congested skies:} Dense formations and rapid maneuvers demand extreme tail-latency and ultra-high reliability
& C2 scheduling/isolation, packet duplication + multi-connectivity, fast congestion control
& Tail latency, C2 freshness (age of information), reliability under blockage \\
\hline
\textbf{Cyber-physical security:} GNSS spoofing, C2 injection, and jamming can trigger immediate physical hazards; mesh routing and edge offloading expand attack surface under tight compute/energy budgets
& Lightweight authentication, physical layer security, anomaly detection
& Detection delay, false alarm and miss, time-to-safe-state \\
\hline
\textbf{Scalability vs.\ battery limits:} More links, sensing, and inference increase energy consumption, and relay imbalance can break mesh connectivity under limited battery capacity
& Energy-aware scheduling/routing, relay-load balancing, trajectory planning, opportunistic charging coordination
& Mission time, energy per control update, supported density. \\
\hline
\textbf{Traffic management \& regulations:} UTM must scale to support massive autonomous operations, while regional policy constraints directly shape network topology and service provisioning
& Hybrid UTM, standardized UTM--LAWN interfaces, policy-aware routing
& Compliance rate, conflict resolution time\\
\hline
\textbf{Explainable \& robust AI:} Black-box decision making under distribution shifts or adversarial inputs undermine trust, while errors can rapidly propagate through multi-agent closed-loop interactions
& Uncertainty-aware inference/planning, XAI summaries/attribution, adversarial robustness and credibility checks
& Calibration, worst-case violations, AI-induced failure rate\\
\hline
\textbf{Intermittent connectivity \& partitioned swarms:} Blockage and mobility can fragment the aerial mesh
& Opportunistic relaying, partition-aware task allocation, redundant safety signaling with local fallback control
& Partition frequency \& duration, service continuity, mission utility. \\
\hline
\end{tabular}
\end{table*}

\section{Cross-layer Challenges of the LAWN} 
In previous sections, we have discussed the four‑plane architecture, enabling technologies, and potential applications of LAWNs. In practice, a LAWN must allocate spectrum, computation, and energy under uncertainty while preserving mission-grade reliability and regulatory compliance. Below, we outline key challenges that can serve as a roadmap for future research and standardization.

\subsection{3D Spectrum Coexistence}
Unlike traditional terrestrial networks, the LAWN operates in 3D space, significantly complicating the spectrum coexistence problem. Low-altitude drones share sub-6 GHz, C-band, and mmWave frequency bands already occupied by 5G and upcoming 6G infrastructure. As drones maneuver vertically and horizontally, their antenna beams continually shift orientation, leading to interference patterns that classic spectrum management frameworks cannot deal with. Addressing this challenge involves characterizing altitude-dependent interference, as drones flying at different heights may cause varying interference impacts on terrestrial users. Towards this, ``dynamic" 3D interference modeling and characterization beyond the existing one in \cite{sharma2019random} are prerequisites, considering the mobility and varying heights of drones as well as their antenna orientations. Furthermore, predictive beam scheduling algorithms must be developed, combining real-time flight mode data (e.g., drone altitude, heading, and velocity) with channel state information, enabling proactive management of interference rather than naive interference mitigation.

\subsection{Time Synchronization and Localization Integrity}
Accurate time synchronization and reliable localization are foundational to many LAWN functions. In practice, however, low-altitude operations frequently take place in GNSS-challenging environments such as urban canyons and post-disaster areas with degraded infrastructure. Under these conditions, clock drift, multipath, and intermittent visibility of satellites can lead to biased position estimates and inconsistent timing across drones. Such errors may propagate through the sensing and control planes, degrading collision-avoidance performance. Moreover, malicious threats such as spoofing can further degrade timing and positioning, making integrity monitoring as important as raw accuracy. Addressing this challenge requires  multi-sensor fusion with network-assisted synchronization over D2D/D2G links, together with integrity measures that can detect and bound localization failures in real time. 

\subsection{Control with Low Latency in Congested Skies}
With the rapid development of LAWN, it is expected that the low-altitude airspace will become increasingly crowded. In such scenarios, drones may fly in close formations, maneuver rapidly to avoid obstacles, raising the demand for end-to-end control loops with extremely stringent latency constraints\cite{zhou2026low}. On the one hand, short frame structures, minimal preambles, and Doppler-resilient waveforms will be adopted for reducing transmission delays. On the other hand, the data plane scheduler should swiftly re-allocate resources for critical C2 message and non-urgent data streams. Developing congestion control algorithms and adaptive network slicing schemes that are capable of dynamically adjusting priorities and resources within microseconds is of great importance.

\subsection{Cyber-Physical Security}
The close integration of cyber and physical components in LAWN poses critical security challenges. Drones in low-altitude airspace are especially vulnerable to attacks such as GNSS spoofing, C2 injection, and communication jamming, which may result in physical threats and impact public safety. To ensure cyber-physical security in LAWN, developing robust and computationally efficient cryptographic protocols is necessary. Moreover, designing onboard-embedded AI algorithms to detect abnormal or malicious behavior can prevent LAWN from severe incidents. In addition, LAWN can harness physical layer security methods, such as artificial-noise beamforming, cooperative jamming, adaptive waveform design {and covert communications}, to prevent eavesdropping and jamming directly at the radio interface with minimal computational overhead.

\subsection{Scalability Versus Battery Limits}
The effectiveness of LAWN is directly tied to its ability to scale up operations. Each additional wireless link, or computational task increases the total energy consumption, leading to a significant challenge given the limited battery capacities of drones. Addressing this challenge requires developing advanced battery technologies with higher energy density and faster recharge rates. Alongside battery advancements,  designing energy-aware scheduling and routing mechanisms is also of great significance. 

\subsection{Resilience to Intermittent Connectivity and Partitioned Swarms}
A defining feature of LAWNs is rapid topology change under mobility and blockage. In cluttered low-altitude environments, LoS links can be abruptly blocked and the aerial mesh may temporarily split into multiple partitions. These intermittent and partitioned conditions challenge conventional routing and scheduling, which often assume persistent end-to-end paths and stable neighbor relations. Robust LAWN operation therefore requires connectivity-aware and disruption-tolerant mechanisms that can gracefully degrade service, such as opportunistic relaying and partition-aware task allocation that avoids assigning tightly coupled tasks to disconnected sub-swarms. 

\subsection{Traffic Management and Global Regulations}
The deployment of LAWN requires effective traffic management and regulatory compliance to ensure safety and operational integrity in low-altitude airspace. However, managing this dynamic 3D environment presents several challenges. First, current traffic management systems were primarily designed for conventional aviation and struggle to accommodate the scale, speed, and autonomy of low-altitude drones. This necessitates new unmanned traffic management (UTM) frameworks that can dynamically handle both cooperative and uncooperative aircraft, maintain separation, and adapt to real-time changes such as weather conditions or communication failures. Furthermore, existing regulations often lag behind technological advancements. There is a lack of harmonized standards for drone registration, remote identification, flight permissions, and data sharing. Future solutions are expected to have hybrid UTM architectures that balance centralized oversight with decentralized autonomy, supported by technologies like blockchain for transparent access control. 

\subsection{Explainable and Robust AI}
Integrating AI into LAWN poses unique challenges due to the inherently black-box nature of deep learning models, raising issues of explainability and robustness\cite{10292755}. While the AI-driven algorithms are powerful, their decision-making processes are typically difficult to interpret, thus undermining trust and practical reliability. Furthermore, such models are often sensitive to adversarial data, making them vulnerable to manipulated inputs designed to trigger incorrect responses. To overcome these challenges, explainable AI (XAI) techniques are critical, enabling visualization and attribution of features, simplification of models, and sensitivity analysis to detect and mitigate adversarial impacts. Incorporating robust AI methods, such as adversarial training and model credibility assessment, ensures that LAWN can reliably support critical tasks even in adversarial conditions.

\section{Conclusion and Outlook}
In this article, we have discussed the concept of LAWN, which offers a new framework for realizing intelligent, resilient, and multifunctional wireless systems in the low-altitude airspace. By tightly integrating data transmission, control, environmental sensing, and intelligence/computing, the LAWN extends the traditional role of drones as communication relays or remote sensors. Instead, it establishes a reconfigurable, three-dimensional infrastructure capable of adapting dynamically to mission objectives, environmental conditions, and airspace constraints. Realizing this vision requires cross-disciplinary collaboration across wireless communications, control theory, robotics, embedded AI, and aviation regulation. Looking ahead, we anticipate that the LAWN will play a foundational role in the emerging low-altitude ecosystems. It has the potential to transform how cities manage logistics, how rural areas access services, and how societies monitor and interact with the environment. This article aims to provide both a conceptual foundation and a research roadmap for advancing LAWN, encouraging the academic and industrial communities to jointly shape the future of autonomous aerial networks.

\bibliographystyle{IEEEtran}
\bibliography{ref}

@misc{dronemarket,
  title = {Global drone market to rise from {USD}28.5b in 2021 to {USD}260b in 2030},
  howpublished = {\url{https://www.unmannedairspace.info/latest-news-and-information}},
  note = {Accessed: 2024-03-06}
}

@ARTICLE{10475844,
  author={Cheng, Yanyu and Niyato, Dusit and Du, Hongyang and Miao, Chunyan and Kim, Dong In},
  journal={IEEE Wireless Commun.}, 
  title={Goal-Oriented Semantic Information Transmission with Message-Sharing {NOMA}}, 
  year={2024},
  volume={31},
  number={3},
  pages={309-315}}

@article{sharma2019random,
  title={Random {3D} mobile {UAV} networks: Mobility modeling and coverage probability},
  author={Sharma, Pankaj K and Kim, Dong In},
  journal={IEEE Transactions on Wireless Communications},
  volume={18},
  number={5},
  pages={2527--2538},
  year={2019},
  publisher={IEEE}
}

@book{zhou2026low,
  title={Low-latency Communication: Theory and Application},
  author={Lin Zhou and Lin Bai},
  year={2026},
  publisher={Springer Nature}
}

@article{zhou2025twentyq,
  title={Twenty Questions with Random Error},
  author={Lin Zhou and Alfred O. Hero},
  journal={Foundations and Trends{\textregistered} in Communications and Information Theory},
  volume={22},
  number={4},
  pages={394--604},
  year={2025},
  publisher={Now Publishers, Inc.}
}

@article{yuan2021integrated,
  title={Integrated sensing and communication-assisted orthogonal time frequency space transmission for vehicular networks},
  author={Yuan, Weijie and Wei, Zhiqiang and Li, Shuangyang and Yuan, Jinhong and Ng, Derrick Wing Kwan},
  journal={IEEE J. Sel. Topic. Signal Process.},
  volume={15},
  number={6},
  pages={1515--1528},
  year={2021},
  publisher={IEEE}
}

@inproceedings{li2021cross,
  title={Cross modal compression: Towards human-comprehensible semantic compression},
  author={Li, Jiguo and Jia, Chuanmin and Zhang, Xinfeng and Ma, Siwei and Gao, Wen},
  booktitle={Proc. ACM Int. Conf. Multimedia},
  pages={4230--4238},
  year={2021}
}

@article{tischler2018system,
  title={System identification methods for aircraft flight control development and validation},
  author={Tischler, Mark B},
  journal={Advances in aircraft flight control},
  pages={35--69},
  year={2018}
}

@article{zeng2019accessing,
  title={Accessing from the sky: A tutorial on {UAV} communications for {5G} and beyond},
  author={Zeng, Yongs and Wu, Qingqing and Zhang, Rui},
  journal={Proc. IEEE},
  volume={107},
  number={12},
  pages={2327--2375},
  year={2019},
  publisher={IEEE}
}

@article{kaleem2019uav,
  title={{UAV}-empowered disaster-resilient edge architecture for delay-sensitive communication},
  author={Kaleem, Zeeshan and Yousaf, Muhammad and Qamar, Aamir and Ahmad, Ayaz and Duong, Trung Q and Choi, Wan and Jamalipour, Abbas},
  journal={IEEE Network},
  volume={33},
  number={6},
  pages={124--132},
  year={2019},
  publisher={IEEE}
}

@article{boubrima2021robust,
  title={Robust environmental sensing using {UAVs}},
  author={Boubrima, Ahmed and Knightly, Edward W},
  journal={ACM Trans. Internet of Things},
  volume={2},
  number={4},
  pages={1--20},
  year={2021},
  publisher={ACM New York, NY, USA}
}

@article{cohen2021urban,
  title={Urban air mobility: History, ecosystem, market potential, and challenges},
  author={Cohen, Adam P and Shaheen, Susan A and Farrar, Emily M},
  journal={IEEE Trans. Int. Transport. Sys.},
  volume={22},
  number={9},
  pages={6074--6087},
  year={2021},
  publisher={IEEE}
}

@article{yuan2023new,
  title={New delay Doppler communication paradigm in 6G era: A survey of orthogonal time frequency space ({OTFS})},
  author={Yuan, Weijie and Li, Shuangyang and Wei, Zhiqiang and Cui, Yuanhao and Jiang, Jiamo and Zhang, Haijun and Fan, Pingzhi},
  journal={China Commun.},
  volume={20},
  number={6},
  pages={1--25},
  year={2023},
  publisher={IEEE}
}

@article{wei2021orthogonal,
  title={Orthogonal time-frequency space modulation: A promising next-generation waveform},
  author={Wei, Zhiqiang and Yuan, Weijie and Li, Shuangyang and Yuan, Jinhong and Bharatula, Ganesh and Hadani, Ronny and Hanzo, Lajos},
  journal={IEEE Wireless Commun.},
  volume={28},
  number={4},
  pages={136--144},
  year={2021},
  publisher={IEEE}
}

@article{liu2022integrated,
  title={Integrated sensing and communications: Toward dual-functional wireless networks for {6G} and beyond},
  author={Liu, Fan and Cui, Yuanhao and Masouros, Christos and Xu, Jie and Han, Tony Xiao and Eldar, Yonina C and Buzzi, Stefano},
  journal={IEEE J. Sel. Areas Commun.},
  volume={40},
  number={6},
  pages={1728--1767},
  year={2022},
  publisher={IEEE}
}

@article{chang2024survey,
  title={A survey on evaluation of large language models},
  author={Chang, Yupeng and Wang, Xu and Wang, Jindong and Wu, Yuan and Yang, Linyi and Zhu, Kaijie and Chen, Hao and Yi, Xiaoyuan and Wang, Cunxiang and Wang, Yidong and others},
  journal={ACM Trans. Intell. Sys. Technol.},
  volume={15},
  number={3},
  pages={1--45},
  year={2024},
  publisher={ACM New York, NY}
}

@article{xie2021uav,
  title={{UAV}-enabled wireless power transfer: A tutorial overview},
  author={Xie, Lifeng and Cao, Xiaowen and Xu, Jie and Zhang, Rui},
  journal={IEEE Trans. Green Commun. Netw.},
  volume={5},
  number={4},
  pages={2042--2064},
  year={2021},
  publisher={IEEE}
}

@article{pupiales2021multi,
  title={Multi-connectivity in mobile networks: Challenges and benefits},
  author={Pupiales, Carlos and Laselva, Daniela and De Coninck, Quentin and Jain, Akshay and Demirkol, Ilker},
  journal={IEEE Commun. Mag.},
  volume={59},
  number={11},
  pages={116--122},
  year={2021},
  publisher={IEEE}
}

@article{zhao2025generative,
  title={Generative {AI}-enabled wireless communications for robust low-altitude economy networking},
  author={Zhao, Changyuan and Wang, Jiacheng and Zhang, Ruichen and Niyato, Dusit and Sun, Geng and Du, Hongyang and Kim, Dong In and Jamalipour, Abbas},
  journal={arXiv preprint arXiv:2502.18118},
  year={2025}
}

@ARTICLE{10040750,
  author={de Souza, João Henrique Inacio and Filho, José Carlos Marinello and Amiri, Abolfazl and Abrão, Taufik},
  journal={IEEE Trans. Veh. Technol.}, 
  title={{QoS}-Aware User Scheduling in Crowded {XL-MIMO} Systems Under Non-Stationary Multi-State {LoS/NLoS} Channels}, 
  year={2023},
  volume={72},
  number={6},
  pages={7639-7652}}

@ARTICLE{10689714,
  author={Duan, Guang-Ren},
  journal={IEEE Tran. Cyber.}, 
  title={Fully Actuated System Approach for Control: An Overview}, 
  year={2024},
  volume={54},
  number={12},
  pages={7285-7306}}

@ARTICLE{10292755,
  author={Khan, Nasir and Coleri, Sinem and Abdallah, Asmaa and Celik, Abdulkadir and Eltawil, Ahmed M.},
  journal={IEEE Commun. Mag.}, 
  title={Explainable and Robust Artificial Intelligence for Trustworthy Resource Management in {6G} Networks}, 
  year={2024},
  volume={62},
  number={4},
  pages={50-56}}

@inproceedings{deng2024drone,
  title={A Drone Delivery Problem Arising in the Meal Delivery based on {VNS} Algorithm},
  author={Deng, Jie and He, Yandong and Gao, Benhe and Liu, Pan},
  booktitle={Proc. 2024 Int. Conf. Control Theory App.},
  pages={123--127},
  year={2024},
  organization={IEEE}
}

@article{perez2023urban,
  title={Urban firefighting drones: Precise throwing from {UAV}},
  author={Perez-Saura, David and Fernandez-Cortizas, Miguel and Perez-Segui, Rafael and Arias-Perez, Pedro and Campoy, Pascual},
  journal={J. Int.  Rob. Sys.},
  volume={108},
  number={4},
  pages={66},
  year={2023},
  publisher={Springer}
}

@article{yang2017unmanned,
  title={Unmanned aerial vehicle remote sensing for field-based crop phenotyping: current status and perspectives},
  author={Yang, Guijun and Liu, Jiangang and Zhao, Chunjiang and Li, Zhenhong and Huang, Yanbo and Yu, Haiyang and Xu, Bo and Yang, Xiaodong and Zhu, Dongmei and Zhang, Xiaoyan and others},
  journal={Frontiers in Plant Science},
  volume={8},
  pages={1111},
  year={2017},
  publisher={Frontiers Media SA}
}

@misc{pge2025,
author = {PG\&E Corporation},
title = {PG\&E's Aerial System Drone Fleet Supports Safe, Reliable Energy System},
year = {2025},
url = {https://investor.pgecorp.com/news-events/press-releases/press-release-details/2025/PGEs-Aerial-System-Drone-Fleet-Supports-Safe-Reliable-Energy-System/default.aspx},
note = {Accessed: December 03, 2025}
}

@article{liu2024design,
  title={Design and implementation for a {UAV}-based streaming media system},
  author={Liu, Zhichao and Jiang, Yi},
  journal={Ad Hoc Networks},
  volume={156},
  pages={103443},
  year={2024},
  publisher={Elsevier}
}

@techreport{3gpp2024service,
author = {{3GPP}},
title = {Service requirements for cyber-physical control applications in vertical domains (Release 18)},
number = {TR 22.125 V18.0.0},
year = {2024},
month = {March}
}

@article{saad2024non,
  title={Non-terrestrial networks: An overview of {3GPP release 17} \& 18},
  author={Saad, Malik Muhammad and Tariq, Muhammad Ashar and Khan, Muhammad Toaha Raza and Kim, Dongkyun},
  journal={IEEE Int. Things Mag.},
  volume={7},
  number={1},
  pages={20--26},
  year={2024},
  publisher={IEEE}
}

\end{document}